\documentclass[12pt,english]{article}
\usepackage[T1]{fontenc}
\usepackage[latin9]{inputenc}
\usepackage{amsmath}
\usepackage{amssymb}
\usepackage{graphicx}
\usepackage[multiple]{footmisc}

\makeatletter

\date{} 

\usepackage{babel}

\usepackage{babel}

\makeatother

\usepackage{babel}

\title{Exploring Spiral Inflation in String Theory}
\begin{document}

\author{Pontus Ahlqvist$^1$%
\thanks{pontus@phys.columbia.edu%
}, Brian Greene$^1$%
\thanks{greene@phys.columbia.edu%
}, and David Kagan$^2$%
\thanks{dkagan@umassd.edu%
}\\
 \\
 $^1$\emph{Institute of Strings, Cosmology, and Astroparticle Physics}\\
 \emph{Department of Physics}\\
 \emph{Columbia University, New York, NY 10027, USA}
 \\
 \\
 $^2$\emph{Department of Physics}\\
 \emph{University of Massachusetts Dartmouth, North Dartmouth, MA 02747, USA}}
 
\maketitle
\begin{abstract}

We investigate the possibility that spiral inflation can be realized using the near-conifold flux potentials for the complex structure moduli  in type IIB string theory compactified on a Calabi-Yau manifold. Using the explicit form of the flux potential for complex structure moduli, we provide analytical and numerical arguments showing that spiral inflation is difficult to support. We also show that for this sector of low energy string theories, a viable spiral inflationary scenario would owe its success to a de Sitter-like vacuum energy, with minimal reliance on the non-gradient flow field trajectories which characterize spiral inflation. We thus conclude that even though the near conifold region has the requisite multi-sheeted potential called for by spiral inflation, generically it appears that spiral inflation is not realized using the complex structure flux potential alone.

\end{abstract}
\newpage
\tableofcontents
\newpage

\section{Introduction}

A novel form of multifield slowroll inflation, called spiral inflation, was proposed in \cite{Spiral}. Rather than relying on the traditional flat potentials of standard slow roll inflation--a characteristic difficult to achieve within the context of string theory--spiral inflation considers fields evolving along trajectories that do not follow the gradient of the potential. Sustained inflation can then be achieved at the price of requiring a multivalued potential. The trajectories correspond to circular orbits in a central potential where each revolution around the origin takes the field onto a different sheet of the potential. As pointed out in \cite{Spiral}, this situation is natural in the context of type IIB string theory compactified on a Calabi-Yau since such multivalued potentials generically occur near the conifold locus of complex structure flux potentials. 

In this paper, we undertake a detailed investigation of the possibility of realizing spiral inflation with flux potentials located in the the near conifold region. Without explicitly stabilizing any additional moduli, we analyze the near conifold potential as a model with adjustable parameters, and argue that for the class of examples one encounters in string theory, spiral inflation requires potentials with a large de Sitter-like vacuum energy at their local minima. Moreover, our numerical simulations show that even in such potentials, the dynamics coaxes the field trajectories to rapidly violate the necessary conditions for spiral inflation. Thus, when sustained inflation is possible, it is due to the de Sitter-like properties of the potential rather than the new field trajectories of spiral inflation itself. For other, less tractable regions, we give heuristic arguments for why we anticipate that spiral inflation is also unlikely to be realized.

\section{Background}
We briefly review the main ideas of spiral inflation and flux potentials in string theory. More detailed analyses can be found in \cite{Spiral} and \cite{Conifunneling}.

\subsection{Spiral Inflation}

As is well known, the generalization of the usual single-field second slow roll condition to multifield inflation is nontrivial, with a variety of distinct but related proposals in the literature \cite{Spiral,TegmarkSecondSlowRoll,BerglundSecondSlowRoll,SasakiSecondSlowRoll}. Although requiring the Hubble parameter to change slowly gives  the standard first slow roll parameter $\epsilon$, requiring $\epsilon$ itself to change slowly leads to possibilities other than the standard second slow roll parameter being small. In particular, by requiring the Hubble parameter to change slowly one must (as usual) have potential energy dominated evolution:

\begin{equation}
|\epsilon| = \left|\frac{\dot{H}}{H^2}\right| \ll 1 \hspace{5mm} \rightarrow \hspace{5mm} \frac{1}{2}\dot{\vec{\phi}}\cdot \dot{\vec{\phi}} \, \ll \, V.
\end{equation}

Then requiring $\epsilon$ to also change slowly one finds

\begin{equation}
\left|\frac{\dot{\epsilon}}{\epsilon H}\right|  \ll 1 \hspace{5mm}\rightarrow\hspace{5mm} |\dot{\vec{\phi}} \cdot \ddot{\vec{\phi}}| \,\ll\, H \dot{\vec{\phi}}\cdot \dot{\vec{\phi}}.
\end{equation}

In the single field case, this directly implies that the acceleration of the field must be small. This then leads, via the equations of motion, to the conclusion that slow roll inflation is only possible in flat potentials. However as is pointed out in \cite{Spiral}, a novel perspective is available in the multifield case. We do not need the acceleration to be small. Instead, it is sufficient to have it orthogonal to the trajectory, i.e. to have fields move along roughly circular orbits. In short, one can slow roll even in steep potentials. 

In order for such spiral inflation to differ significantly from regular slow roll inflation, one wants a large centripetal acceleration. However, this means that the angular motion must be rapid and so a full revolution will pass quickly. For sustained inflation of this sort, therefore, we need a multivalued potential \cite{Spiral}. Fortunately such potentials are ubiquitous near the conifold locus of complex structure moduli spaces in string theory. A natural suggestion, then, is that one should look for sustained spiral inflation in these regions of moduli space. 

In order for the dot product $\dot{\vec{\phi}} \cdot \ddot{\vec{\phi}}$ to be small, the acceleration in the direction of motion must be negligible. Using the equations of motion for each field,

\begin{equation}
\ddot{\phi}_{i} + 3H\dot{\phi}_{i} + \partial_{i}V = 0,
\end{equation}
it's clear that the Hubble friction must then balance the tilt of the potential in the direction of motion. If we move to polar coordinates $(r,\theta)$ where the direction of motion is along the angular direction, this implies that

\begin{eqnarray}
\dot{r} &=& 0 \nonumber \\
3 H r^2 \dot{\theta} + \frac{\partial V}{\partial \theta} &=& 0.
\end{eqnarray}

An example of a potential that allows such an orbit was given in \cite{Spiral}:

\begin{equation}
V(r,\theta) = V_0 + c \theta + \frac{c^2}{9 \alpha H^2}\frac{r^\alpha}{R^{\alpha+2}}.
\label{SpiralPotential}
\end{equation}

Balancing the Hubble friction fixes the angular velocity (or rather the angular momentum):

\begin{equation}
3 H r^2 \dot{\theta} + \frac{\partial V}{\partial \theta} = 0 \hspace{5mm} \rightarrow \hspace{5mm} \dot{\theta} = -\frac{c}{3 H r^2}.
\end{equation}

Then, looking for a stable circular orbit involves finding the minimum of the effective potential

\begin{equation}
\frac{\partial V_{eff}}{\partial r} = 0 \hspace{5mm} \rightarrow \hspace{5mm} r = R.
\end{equation}

One benefit of balancing the angular tilt against the Hubble friction is that the angular momentum becomes a conserved quantity. This in turn implies that the centrifugal uplift of the effective potential doesn't change as the system evolves. As a result, the initial circular orbit can remain stable for an extended period. 

All of this, of course, only enforces the orthogonality of the acceleration and the velocity. In addition, one must also ensure that the motion is dominated by potential energy. Since the orbit is circular, the kinetic energy is completely due to the angular motion

\begin{equation}
\frac{L^2}{2r^2} = \left(\frac{c}{3 H}\right)^2\frac{1}{2 R^2} = \frac{1}{2}\frac{c^2}{9 H^2}\frac{1}{R^2}.
\end{equation}

Notice that, up to a factor of $\alpha/2$, this is precisely the value of the $r$ dependent part of the potential in equation (\ref{SpiralPotential}). Therefore, in order to simultaneously be potential energy dominated, we must make sure to either choose $V_0$ large or choose initial conditions with a large enough $\theta$ so that the potential is dominated by the $r$ independent parts. This phenomenon is what will present us with a tradeoff for potentials motivated by string theory: in these cases, spiral inflation can only be realized where de Sitter like inflation was already possible.

\subsection{The Scalar Potential in String Theory}

In the low energy supergravity limit of type IIB string theory compactified on a Calabi-Yau manifold, scalar moduli fields are generally abundant, and fall into two classes, K\"ahler and complex strucuture moduli. We focus on the latter fields, which parametrize the smooth ways in which the complex structure of the internal manifold can be deformed while retaining the Calabi-Yau structure. Turning on 3-form fluxes in the internal manifold stabilizes the values of these fields and gives them a mass. We will for simplicity work within the family of single complex modulus models although our results will be quite general since the conifold locus will always be a co-dimension one structure. The resulting scalar potential takes the standard no-scale form

\begin{equation}
 V(\xi,\tau)=\frac{e^{K}}{2\tau_{I}}\left(K^{\xi\bar{\xi}}\left|D_{\xi}W\right|^{2}+K^{\tau\bar{\tau}}\left|D_{\tau}W\right|^{2}\right).
\label{scalarPotential}
\end{equation}

Here, $K$ is the K\"ahler potential, $K_{\xi\bar{\xi}}$ is the K\"ahler metric, and $W$ is the Gukov-Vafa-Witten superpotential given by

\begin{equation}
W = (\vec{\mathcal{F}}-\tau \vec{\mathcal{H}})\cdot \vec{\Pi}(\xi)
\end{equation}

and we compute covariant derivatives according to

\begin{equation}
D_{\xi}W = W_{\xi} + K_\xi W.
\end{equation}

In these expressions, $\tau$ is the axio-dilaton while $\xi$ is the complex structure modulus. Also, $\vec{\mathcal{F}}=(\mathcal{F}_0,\mathcal{F}_1,\mathcal{F}_2,\mathcal{F}_3)$ and $\vec{\mathcal{H}}=(\mathcal{H}_0,\mathcal{H}_1,\mathcal{H}_2,\mathcal{H}_3)$ are the components of the 3-form fluxes $\mathcal{F}$ and $\mathcal{H}$ in a certain basis of 3-forms and $\Pi(\xi) = (\Pi_3,\Pi_2,\Pi_1,\Pi_0)$ is a vector of periods of the holomorphic 3-form $\Omega_3$ along the dual cycles. We set the component $\mathcal{H}_3=0$ using modular invariance. We will take $\xi=0$ as the conifold point and $\Pi_3(\xi)$ as the period along the collapsing cycle. We normalize $\xi$ so that near the conifold, $\Pi_3(\xi) = \xi$. 

\section{Expanding the potential}

As our goal is to analyze the possibility of inflation in the near conifold region, we begin by expanding the potential and K\"ahler metric to leading order in $r = |\xi|$. We will assume that we are sufficiently near the conifold to neglect any angular dependence in the K\"ahler metric. In other words since any angular dependence in the K\"ahler metric comes from linear terms, we suppose that $|\xi| \ll 1$. 

\subsection{K\"ahler Metric, Connection, and Potential}

Letting $\Pi(\xi)$ denote the vector of periods and $\xi$ denote
the complex modulus that vanishes at the conifold point, the unwarped
K\"ahler potential is
\begin{equation}
K^{o}=-\log\left(-i\Pi^{\dagger}Q\Pi\right)
\end{equation}
where $Q$ is the symplectic form
\begin{equation}
Q=\left(\begin{array}{cccc}
0 & 0 & 0 & -1\\
0 & 0 & 1 & 0\\
0 & -1 & 0 & 0\\
1 & 0 & 0 & 0
\end{array}\right).
\end{equation}
The warp corrected K\"ahler potential (derived by integrating back from
the warp corrections to the K\"ahler metric) is given by
\begin{equation}
K=K^{o}+\widehat{K}=K^{o}+9C_{w}\left|\xi\right|^{2/3}.
\end{equation}

The warping corrections to the K\"ahler connection and metric are then given by

\begin{align*}
K_{\xi} & =K_{\xi}^{o}+\widehat{K}_{\xi}=K_{\xi}^{o}+3C_{w}\frac{\bar{\xi}^{1/3}}{\xi^{2/3}}\\
K_{\xi\bar{\xi}} & =K_{\xi\bar{\xi}}^{o}+\widehat{K}_{\xi\bar{\xi}}=K_{\xi\bar{\xi}}^{o}+C_{w}\left|\xi\right|^{-4/3}.
\end{align*}

The expansion of the unwarped K\"ahler potential and its derivatives
near the conifold points then becomes:
\begin{eqnarray}
K^{o} &=&-\log\left(k+\frac{\left|\xi\right|^{2}}{2\pi}\log\left|\xi\right|^{2}\right)\to-\log k \nonumber\\
K_{\xi}^{o} &=&-e^{K^{o}}\left(k_{\xi}+\frac{\bar{\xi}}{2\pi}\left(\log\left|\xi\right|^{2}+1\right)\right)\to-\frac{k_{\xi}}{k}\nonumber\\
K_{\xi\bar{\xi}}^{o} &\to& \left(\left|K_{\xi}^{o}\right|^{2}-\frac{1}{\pi k}-\frac{k_{\xi\bar{\xi}}}{k}\right)-\frac{1}{2\pi k}\log\left|\xi\right|^{2}\to\kappa-\frac{1}{2\pi k}\log\left|\xi\right|^{2}
\end{eqnarray}
where we've introduced the regular functions $k,\ k_{\xi}$, and $\kappa$.
These approach constant values near the conifold point. In fact since we will be expanding everything to lowest nontrivial order in $|\xi|$, we will always evaluate $\kappa, k, k_{\xi}$ as well as other analytic functions at $\xi=0$. The forms that we will use for the various K\"ahler quantities are given as (note that the warping correction is only relevant for $K_{\xi}$ and $K_{\xi \bar{\xi}}$)

\begin{eqnarray}
e^{K(r,\theta)} &=& \frac{1}{k} \nonumber \\
K_{\xi}(r,\theta) &=& -\frac{k_{\xi}}{k} + 3 C_{w} r^{-1/3} e^{-i\theta} \nonumber \\
K_{\xi \bar{\xi}}(r,\theta) &=& \kappa - \frac{1}{\pi k} \log r + C_w r^{-4/3}.
\label{K\"ahlerQuantities}
\end{eqnarray}

Here we use polar notation $\xi = r e^{i\theta}$. Again, all the functions $k,k_{\xi},$ etc. should be evaluated at $\xi = 0$ (since their $\xi$ dependence is subleading) and are therefore simply model dependent constants. Note that the angular dependence for the metric would come from the terms in $\kappa$ that are linear in $|\xi|$ which is what motivates us to expand everything to lowest order in $r$,  as noted above.

\subsection{Superpotential and its Covariant Derivative}

The superpotential $W=\left(\mathcal{F}-\tau\mathcal{H}\right)\cdot\Pi$, is a regular function that goes to a constant near the conifold point (note that ``regular'' includes terms like $\xi\log\xi$). The ordinary derivative is given by 

\begin{equation}
W_{\xi}=w_{\xi}+\frac{\mathcal{F}_{3}}{2\pi i}\log\xi
\end{equation} 
where $\mathcal{F}_{3}$ and $\mathcal{H}_{3}=0$ are the fluxes piercing the shrinking cycle (we've restricted to this particular case for simplicity). We have also lumped all of the regular terms into $w_{\xi}$. Notice that $W_\xi$ diverges due to the logarithm. The full  covariant derivative along with the warp correction is then 

\begin{equation}
D_{\xi}W = D^{o}_{\xi}W+3C_{w}\frac{\bar{\xi}^{1/3}}{\xi^{2/3}}W\to\left(w_{\xi}-\frac{k_{\xi}}{k}w\right)+\left(\frac{\mathcal{F}_{3}}{2\pi i}\log\xi+3C_{w}\frac{\bar{\xi}^{1/3}}{\xi^{2/3}}w\right).
\end{equation}

For simplicity we lump all of the regular terms into $\omega$ so that to this order, 

\begin{equation}
D_{\xi}W=\omega+\frac{\mathcal{F}_{3}}{2\pi i}\log\xi+3C_{w}\frac{\bar{\xi}^{1/3}}{\xi^{2/3}}w.
\end{equation}

The covariant derivative in the axio-dilaton direction is
\begin{equation}
D_{\tau}W=\frac{1}{\tau-\overline{\tau}}\left(\mathcal{F}-\overline{\tau}\mathcal{H}\right)\cdot\Pi=\frac{1}{\tau-\overline{\tau}}M.
\end{equation}
Here we have defined the short hand notation $M = \left(\mathcal{F}-\overline{\tau}\mathcal{H}\right)\cdot\Pi$ which, since it is also regular at $\xi = 0$, can be thought of as constant in $\xi$ (although it of course still depends on $\tau$). In terms of the coordinates $r,\theta$ we write

\begin{equation}
D_{\xi}W = \omega + \frac{\mathcal{F}_{3}}{2\pi i}\log r + \frac{\mathcal{F}_{3}}{2\pi }\theta+3C_{w}r^{-1/3}e^{-i\theta} w.
\label{DW}
\end{equation}

\subsection{Scalar Potential}

To put everything together to determine the near conifold form for the potential, we use polar coordinates and write $\omega = |\omega|e^{i\theta_\omega}$ and $w = |w|e^{i\theta_w}$, yielding


\begin{eqnarray}
|D_{\xi}W|^2 &=& |\omega|^2+\left(\frac{\mathcal{F}_3}{2\pi}\right)^2(\log r)^2+\left(\frac{\mathcal{F}_3}{2\pi}\right)^2\theta^2 + 9C_w^2r^{-2/3}|w|^2+\frac{\mathcal{F}_3}{\pi}|\omega|\sin(\theta_\omega)\log r \nonumber \\
&& +\frac{\mathcal{F}_3}{\pi}|\omega|\cos(\theta_\omega)\theta +6C_w |\omega||w|r^{-1/3} \cos(\theta_\omega+\theta_w-\theta)+ \nonumber \\
&& \frac{3\mathcal{F}_3}{\pi}|w|C_w r^{-1/3}\log r \sin(\theta_w-\theta)+ \frac{3\mathcal{F}_3}{\pi} |w|C_w
r^{-1/3}\theta\cos(\theta_w-\theta).
\label{DWExpansion}
\end{eqnarray}

In comparing with the model potential presented in \cite{Spiral} and also repeated in equation (\ref{SpiralPotential}) we see that the most natural way to replicate spiral inflation would be to focus on regimes where the potential has a simple non-periodic angular dependence. This is achieved by taking $r \gg C_w^3$ so that the terms with periodic dependence in $\theta$ become subleading. This amounts to neglecting the warping corrections in the K\"ahler connection. In this regime, the above expression simplifies as

\begin{equation}
|D_{\xi}W|^2 = |\omega|^2+\left(\frac{\mathcal{F}_3}{2\pi}\right)^2(\log r)^2+\left(\frac{\mathcal{F}_3}{2\pi}\right)^2\theta^2+\frac{\mathcal{F}_3}{\pi}\omega_I\log r+\frac{\mathcal{F}_3}{\pi}\omega_R\theta.
\end{equation}

We have here reverted to Cartesian form with $\omega = \omega_R+i\omega_I$. We can absorb the terms linear in $\theta$ by redefining the angular variable as $\phi = \theta + 2\pi \omega_R /\mathcal{F}_3$. The full scalar potential then becomes (also taking into account the contribution from the axio-dilaton)

\begin{equation}
V(r,\phi) = V_{0}(r)+V_{1}(r)\phi^2,
\label{potential}
\end{equation}

where we have defined the two $r$-dependent functions $V_0$ and $V_1$ as

\begin{eqnarray}
V_{0}(r) &=& \alpha + \frac{\beta}{K_{\xi\bar{\xi}}(r)} (\log r +\gamma)^2 \nonumber \\
V_{1}(r) &=& \frac{\beta}{K_{\xi\bar{\xi}}(r)}.
\label{V0V1}
\end{eqnarray}

The constants $\alpha,\beta,\gamma$ are given by

\begin{eqnarray}
\alpha &=& \frac{|M|^2}{2 k \tau_I} \nonumber \\
\beta &=& \frac{1}{2 k \tau_I} \left(\frac{\mathcal{F}_3}{2\pi}\right)^2 \nonumber \\
\gamma &=& \frac{2\pi \omega_I}{\mathcal{F}_3}
\end{eqnarray}

\section{Spiraling}

As spiral inflation utilizes field trajectories that are approximately circular, to obtain a sufficiently large number of e-foldings the field must continue its motion onto different sheets thereby requiring a multivalued potential such as the one found near the conifold point $r=0$ in equation (\ref{potential}). To maximize the duration of the spiraling motion, one should initialize the fields at or very near the minimum of the effective potential in the radial direction\footnote{For clarity here, we neglect the influence of the K\"ahler metric, an approximation we will remedy shortly.},

\begin{equation}
V_{eff}(r,\phi) = \frac{L^2}{2r^2}+V(r,\phi).
\end{equation}

 By also balancing the Hubble friction against the tilt in the angular direction, we ensure that angular momentum, and therefore also the centrifugal uplift in the effective radial potential, is conserved. This allows for a sustained circular orbit. 

\subsection{Spiral Inflation vs. de Sitter Space}

Minimizing the effective potential gives

\begin{equation}
\frac{L^2}{r^3} = \frac{\partial V}{\partial r}.
\end{equation}

In terms of the kinetic energy of the field, this implies that 

\begin{equation}
\frac{L^2}{2r^2} = \frac{1}{2}r\frac{\partial V}{\partial r}.
\end{equation}

In order to spiral at the same time as having potential energy dominated dynamics, one must initialize the field in a place where this kinetic energy is much smaller than the potential at that point

\begin{equation}
\frac{1}{2}r\frac{\partial V}{\partial r} \ll V(r,\phi).
\label{SpiralConstraint}
\end{equation}

Now, for any function $f(r)$ that's of the form 

\begin{equation}
f(r) = r^n (\log r)^m
\label{polylogarithmic}
\end{equation}
one sees that $r\partial f/\partial r$ is at least on the same order as $f$ unless $n$ is very small. Furthermore, the radially dependent part of the potential that was suggested in \cite{Spiral} and reproduced in equation (\ref{SpiralPotential}) was precisely of this form for $m=0$. As a result, the only way that the constraint in equation (\ref{SpiralConstraint}) can be satisfied for that potential is if the potential is dominated by the terms independent of $r$. In \cite{Spiral} this was accomplished by adding an arbitrary constant $V_0$ and/or moving sufficiently far up the monodromy ladder (i.e. taking $\phi$ large). As we will see shortly, when this approach is applied to the string inspired potentials above, we find that the support for spiral inflation derives essentially from ordinary de Sitter inflation. But first, let's determine the modifications that arise when a radially dependent K\"ahler metric is introduced.

One can still define an angular momentum $L = K_{\xi\bar{\xi}}(r)r^2\dot{\phi}$ and effective potential

\begin{equation}
V_{eff}(r,\phi) = \frac{L^2}{2 K_{\xi\bar{\xi}}(r) r^2} + V(r,\phi).
\end{equation}

A circular orbit can then be realized when

\begin{equation}
\frac{L^2}{2 K_{\xi\bar{\xi}}(r) r^2}\frac{\frac{\partial }{\partial r}(K_{\xi\bar{\xi}}(r) r^2)}{K_{\xi\bar{\xi}}(r) r^2} = \frac{\partial V}{\partial r}.
\label{circularBalance}
\end{equation}

That is, the kinetic energy must necessarily equal

\begin{equation}
\frac{L^2}{2 K_{\xi\bar{\xi}}(r) r^2} = \frac{K_{\xi\bar{\xi}}(r) r^2}{\frac{\partial }{\partial r}(K_{\xi\bar{\xi}}(r) r^2)}\frac{\partial V}{\partial r}.
\label{kineticEnergy}
\end{equation}

Given the form of the K\"ahler metric from equation (\ref{K\"ahlerQuantities}), the combination $K_{\xi\bar{\xi}(r)} r^2$ consists of terms precisely of the form of equation (\ref{polylogarithmic}). As a result, the right hand side of equation (\ref{kineticEnergy}) is on the same order as $r\partial V/\partial r$ which, because our potential consists of terms of the form displayed in equation (\ref{polylogarithmic}), is also on the same order as the $r$ dependent part of $V$. Requiring the potential to dominate the kinetic energy as the field spirals,  again implies that the potential must be dominated by the terms that are independent of $r$. However, unlike what we found earlier, moving up the monodromy ladder won't have impact since the angular dependent term is not independent of $r$. In fact, the only term in equation (\ref{potential}) that's independent of $r$ is the constant term, $\alpha$. 

Thus, we can only satisfy all the relevant conditions if we take $V\approx \alpha$. Since our potential is essentially bounded from below (unlike the one presented in \cite{Spiral}), we thus find ourselves in de Sitter space. In these flux potentials, then, spiral inflation derives from de Sitter inflation\footnote{Again, de Sitter inlfation here is to be interpreted as inflation driven by contributions to the potential from other fields that may or may not have been fixed.}\footnote{Note that the false vacuum is located at $r=0$, $\phi=0$. However, in this regime many of the terms neglected in equation (\ref{DWExpansion}) become important. Fortunately, even with these terms included, $r=0$ remains a minimum of the potential and as a result we can still conclude that we are in de Sitter space.}, and so is not particularly interesting. 

Nevertheless, it is still important to investigate whether spiral inflation can be sustained in this family of potentials, which we now consider through numerical simulations.  

\section{Numerical Simulations}

Since spiral inflation relies on a nearly circular orbit, we want to initialize our field with $\dot{r}=0$. In order to sustain spiral inflation we also want $\dot{r}$ to remain zero. This is accomplished by initializing the field near the minimum of the effective potential as described in equation (\ref{circularBalance}). As the field evolves, we want to retain this balance, which means that we want to conserve angular momentum. As explained, this is accomplished by setting $\ddot{\phi}=0$ and balancing the Hubble friction against the potential tilt in the angular direction

\begin{equation}
3 H L + \frac{\partial V}{\partial \phi} = 0 \hspace{5mm} \rightarrow \hspace{5mm} \phi = -\frac{3HL}{2V_{1}(r)} = -\frac{3HL}{2\beta}K_{\xi\bar{\xi}}(r).
\label{torqueBalance}
\end{equation}

The sign simply tells us that we need to move down the spiral. We therefore have a one parameter set of initial conditions parametrized by the angular momentum present in the system. Without loss of generality, we will take $\phi$ positive so that $L$ is negative. Once $L$ is specified, the two constraints in equations (\ref{circularBalance}) and (\ref{torqueBalance}) determine the initial location of the field. The radial velocity should be set to zero and the angular velocity is calculable from the angular momentum. This provides a test of whether spiral inflation can be realized: scan through different values of $L$, initialize the field at the optimal location/velocity and let the system evolve. By tracking the value of the dot product between the acceleration and the velocity, we can determine when spiral inflation fails.

\subsection{Initial Conditions}

The optimal initial location for a given $L$ (which we take negative for definiteness), is given by the simultaneous solution of equations (\ref{circularBalance}) and (\ref{torqueBalance}). Using the latter equation in the former gives us 

\begin{equation}
L^2 = \frac{2K_{\xi\bar{\xi}}(r)^2 r^4}{K_{\xi\bar{\xi}}'(r)r^2+2K_{\xi\bar{\xi}}(r)r}\left(-\frac{\beta K_{\xi\bar{\xi}}'(r)}{K_{\xi\bar{\xi}}(r)^2}(\log r+\gamma)^2 + \frac{2\beta}{K_{\xi\bar{\xi}}(r)r}(\log r+\gamma) -\beta K_{\xi\bar{\xi}}'(r)\left(\frac{3 H L}{2\beta}\right)^2\right).
\end{equation}

\begin{figure}[h]
\begin{center}
\includegraphics[width=70mm,natwidth=800,natheight=499]{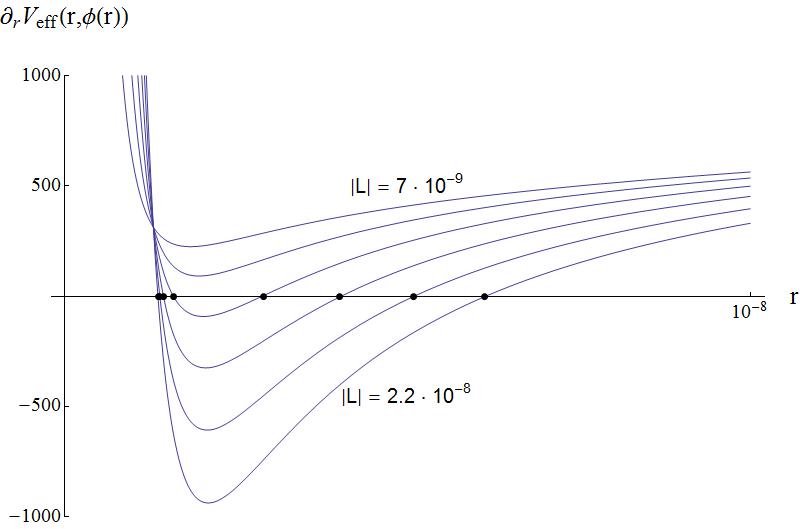}\includegraphics[width=70mm,natwidth=800,natheight=499]{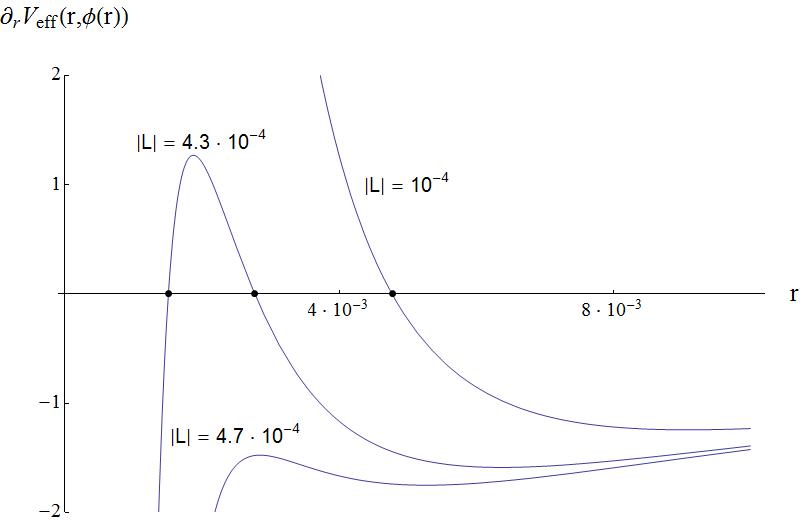}
\end{center}
\caption{$\partial_r V_{eff}$ (evaluated with $\phi=\phi(r)$ where equation (\ref{torqueBalance}) is satisfied) is here plotted for $C_w = 10^{-5}$. (Left) Two additional solutions appear near the conifold point for sufficiently large $L$. (Right) One of these solutions moves out to annihilate another one farther out as the angular momentum is increased.}
\label{dVeffPlot}
\end{figure}

The initial velocities are then given as $\dot{r}=0$ and $\dot{\phi} = L/(K_{\xi\bar{\xi}}(r_0)r_0^2)$ where $r_0$ is the initial radial location. 
Rather than attempting to obtain an analytic solution to the above equation for $r$, we proceed numerically. We generally find that for sufficiently small values for $|L|$, there is a single solution to this equation. As one increases the value of $|L|$, this root remains roughly fixed while two additional roots appear closer to the conifold point. These two new roots then move apart, with one moving closer to the conifold (approaching a limit point) while the other moves away from the conifold point and eventually annihilates the original root, see figure \ref{dVeffPlot}. Since all three of these are legitimate starting locations, we numerically examine each in the next section.

\subsection{Duration of Spiral Inflation}

With the initialization of the field now understood, all that remains is to numerically evolve the system. Such simulations show that the axio-dilaton doesn't evolve in any dramatic way so for simplicity we fix it at $\tau=2i$ for the duration of these simulations. The relevant equations of motion are

\begin{eqnarray}
K_{\xi\bar{\xi}}(r) r^2 \ddot{\phi}+K_{\xi\bar{\xi}}'(r)r^2 \dot{r}\dot{\phi}+2 K_{\xi\bar{\xi}}(r)r \dot{r}\dot{\phi}+3HK_{\xi\bar{\xi}}(r)r^2\dot{\phi}+\frac{\partial V}{\partial \phi} &=& 0\\
K_{\xi\bar{\xi}}(r)\ddot{r} + \frac{1}{2}K_{\xi\bar{\xi}}'(r)\dot{r}^2 + 3 H K_{\xi\bar{\xi}}(r)\dot{r} - \left(\frac{1}{2}K_{\xi\bar{\xi}}'(r) r^2 + K_{\xi\bar{\xi}}(r) r \right)\dot{\phi}^2 + \frac{\partial V}{\partial r} &=& 0.
\end{eqnarray}

We track the evolution of the initial orthogonality condition between the acceleration and the velocity or, more precisely, how fast the kinetic energy is changing. Since our system has a non-canonical K\"ahler metric, the relevant quantity to track is

\begin{equation}
\eta = \frac{1}{K_{\xi\bar{\xi}}(r)(\dot{r}^2 + r^2\dot{\phi}^2)}\frac{d}{dt}\left(K_{\xi\bar{\xi}}(r)(\dot{r}^2 + r^2\dot{\phi}^2)\right).
\label{spiralCondition}
\end{equation}

\begin{figure}[h]
\begin{center}
\includegraphics[width=90mm,natwidth=800,natheight=482]{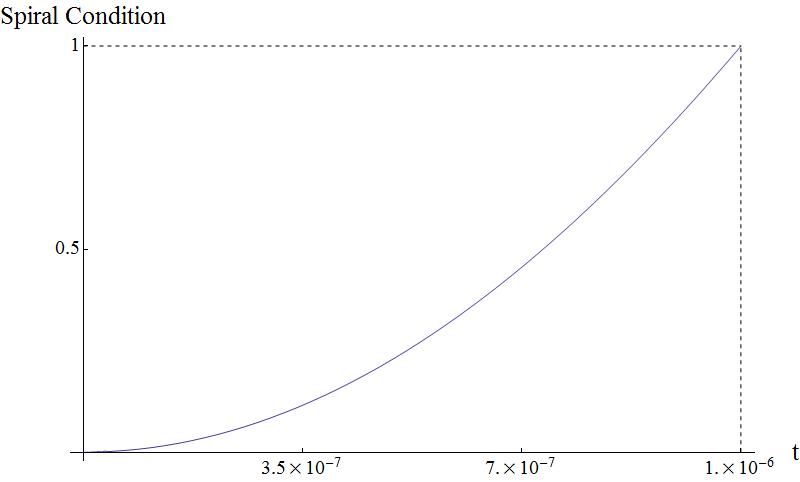}
\end{center}
\caption{Evolution of the absolute value of the spiral condition from equation (\ref{spiralCondition}) plotted here for $C_w=10^{-5}$ and $L=10^{-5}$. This is for the solution nearest the conifold point.}
\label{spiralConditionFig}
\end{figure}

\begin{figure}[h]
\begin{center}
\includegraphics[width=90mm,natwidth=800,natheight=501]{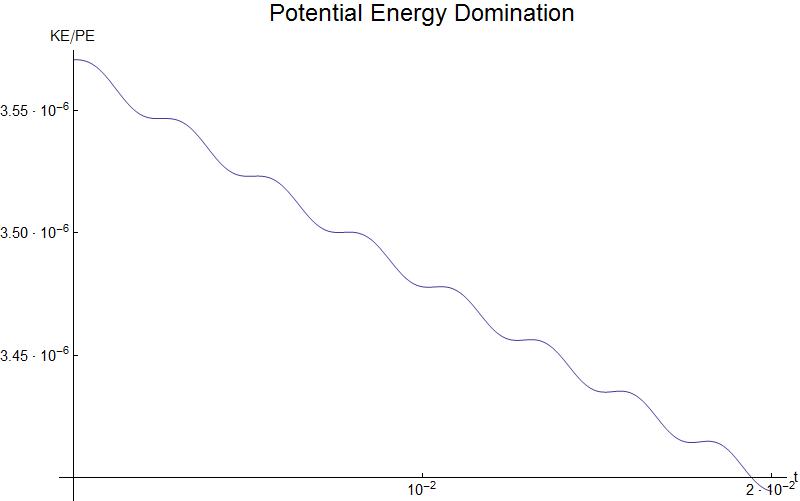}
\end{center}
\caption{Even though the spiral condition in this particular simulation failed at $t=2.2\cdot 10^{-4}$, the motion continues to be potential energy dominated for much longer. This reflects the fact that we have a de Sitter like potential, and is not a result of spiral inflation.}
\label{PotentialDominated}
\end{figure}

\begin{figure}
\begin{center}
\includegraphics[width=90mm,natwidth=800,natheight=513]{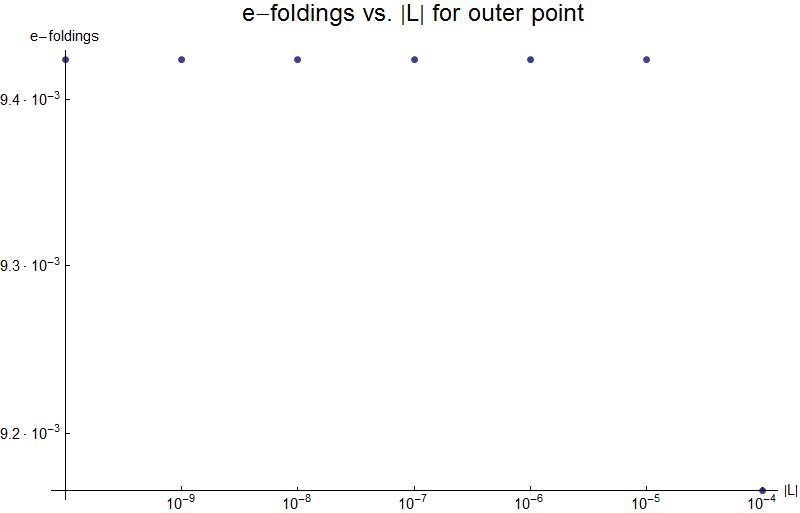}\vspace{10mm}
\includegraphics[width=90mm,natwidth=800,natheight=520]{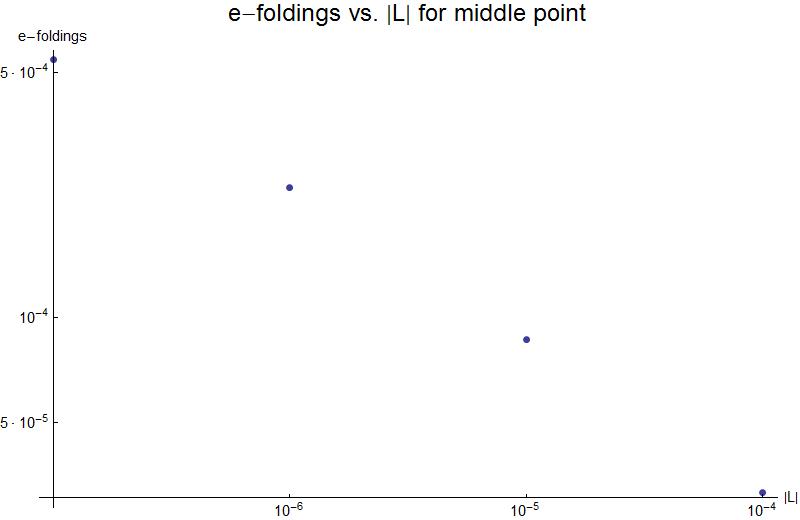}\vspace{10mm}
\includegraphics[width=90mm,natwidth=800,natheight=529]{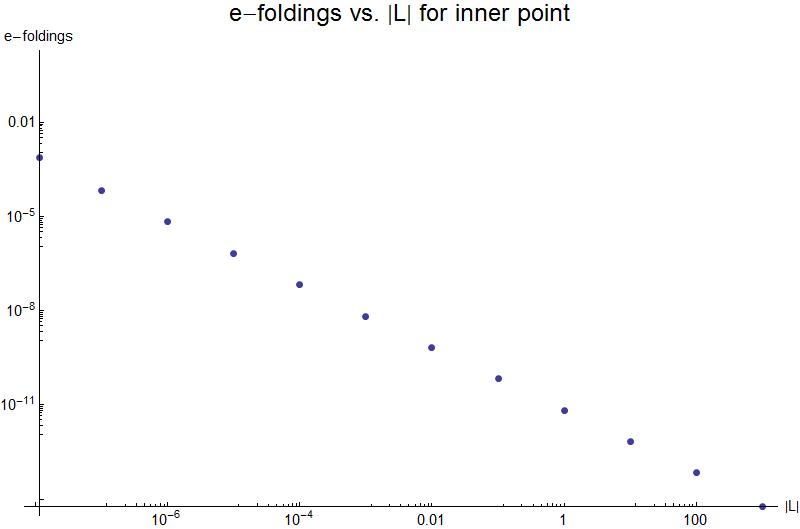}
\end{center}
\caption{Number of e-foldings for various choices of angular momentum $|L|$ for the three possible solutions. Note that the various solutions only exist for a range of $|L|$ (see figure \ref{dVeffPlot})}
\label{efoldingsVsL}
\end{figure}

We do this for a range of angular momenta $10^{-10}<|L|<10^{3}$ and for each of the potential initial conditions. We generically find that the spiral condition, although satisfied at $t=0$, is violated almost immediately. In particular, in figure \ref{spiralConditionFig} the generic time dependence of the spiral condition is displayed. Furthermore, figure \ref{efoldingsVsL} displays the number of e-foldings one obtains for various choices of $|L|$, for various initial locations. As is clear from this analysis, the spiral condition fails almost immediately regardless of the choice of $|L|$. This trend is repeated for various values of $C_w$ although here we only present the data for $C_w = 10^{-5}$. One finds that even though the spiral condition fails almost immediately, the motion continues to be potential energy dominated, see figure \ref{PotentialDominated}, reflecting the fact that the field is rolling inside a de Sitter like region. 

We also note that the initial conditions necessary to initiate spiral inflation generically require us to invoke large values of the angular variable $\phi$. Viewing these monodromies as changes in flux, it's then clear that these large values of $\phi$ equate to large values of fluxes. This inevitably means that we can no longer trust our low energy supergravity model. Nevertheless, for completeness we have still considered these regimes.

\newpage
\newpage
\newpage

\section{Discussion and Conclusions}

We have investigated the possible realization of spiral inflation near the conifold locus of type IIB compactified string models. We have not attempted to fix all moduli and as a result the potentials we study contain various adjustable parameters. We find that spiral inflation requires us to adjust these parameters in such a way as to essentially yield a de Sitter like potential, in which case spiral inflation becomes intertwined with standard de Sitter space and chaotic inflation. In addition to these analytical arguments, we also undertook a numerical study of spiral inflation and showed that rather generically the spiral condition (orthogonality of acceleration and velocity of the field) fails almost immediately for a large range of angular momenta. It thus seems difficult to realize spiral inflation in this variety of complex structure flux potentials.

\section{Acknowledgments}
We thank I-Sheng Yang for helpful suggestions and discussions. This work is supported in part by DOE grant DE-FG02-92ER40699, FQXi grant RFP1-06-19, and STARS grant CHAPU G2009-30 7557.

\end{document}